# Bayesian Co-navigation: Dynamic Designing of the Materials Digital Twins via Active Learning

Boris N. Slautin[1]*, Yongtao Liu[2], Hiroshi Funakubo[3], Rama K. Vasudevan[2], Maxim A. Ziatdinov[4], and Sergei V. Kalinin[4,5]*

[1] Independent researcher, Belgrade, Serbia
[2] Center for Nanophase Materials Sciences, Oak Ridge National Laboratory, Oak Ridge, TN, United States
[3] Department of Material Science and Engineering, Tokyo Institute of Technology, Yokohama, Japan
[4] Pacific Northwest National Laboratory, Richland, WA, United States
[5] Department of Materials Science and Engineering, University of Tennessee, Knoxville, TN, United States
* Corresponding authors: bslautin@gmail.com, sergei2@utk.edu

**Abstract**

Scientific advancement is universally based on the dynamic interplay between theoretical insights, modelling, and experimental discoveries. However, this feedback loop is often slow, including delayed community interactions and the gradual integration of experimental data into theoretical frameworks. This challenge is particularly exacerbated in domains dealing with high-dimensional object spaces, such as molecules and complex microstructures. Hence, the integration of theory within automated and autonomous experimental setups, or theory in the loop automated experiment, is emerging as a crucial objective for accelerating scientific research. The critical aspect is not only to use theory but also on-the-fly theory updates during the experiment. Here, we introduce a method for integrating theory into the loop through Bayesian co-navigation of theoretical model space and experimentation. Our approach leverages the concurrent development of surrogate models for both simulation and experimental domains at the rates determined by latencies and costs of experiments and computation, alongside the adjustment of control parameters within theoretical models to minimize epistemic uncertainty over the experimental object spaces. This methodology facilitates the creation of digital twins of material structures, encompassing both the surrogate model of behavior that includes the correlative part and the theoretical model itself. While demonstrated here within the context of functional responses in ferroelectric materials, our approach holds promise for broader applications, the exploration of optical properties in nanoclusters, microstructure-dependent properties in complex materials, and properties of molecular systems. The analysis code that supports the funding is publicly available at https://github.com/Slautin/2024_Co-navigation/tree/main

# 1. Introduction

Throughout history, one of the central objectives of science has been to uncover fundamental mechanisms and express them through theoretical models to describe physical phenomena. A precise theoretical description not only expands our comprehension of underlying processes but also provides invaluable avenues for optimizing system functionalities. Scientific progress universally relies on the dynamic interplay between theoretical insights, modeling, and experimental discoveries. This holds across various realms of knowledge, irrespective of a system's inherent nature, size, or complexity. At the same time, the integration of theoretical description and experimental results is a slow process, that is traditionally heavily driven by community interaction within the scientific field or requires specific efforts to incorporate experimental data into theoretical models.

The emergence of powerful computing coupled with the development of artificial intelligence (AI) approaches has revolutionized the capabilities for experimental results analysis and modeling complex systems [1–6]. The progress is also supported by the rapid improvement of the data infrastructure bringing abilities to remote operation, co-orchestration of the experiments, and data storing and managing across multiple labs [4,7–11]. The integration of machine learning (ML) approaches with theoretical models has given rise to physics-informed ML (PIML) [12]. PIML leverages knowledge about physics principles to enhance the learning process and improve model performance, making them valuable tools for addressing complex problems. Although physics-informed ML is a relatively new domain area, it has already proven its efficiency in a wide range of various applications [13]. Furthermore, advancements in AI approaches have ushered in a new era for experimental investigations with the automation of synthesis [14–20], as well as the development of automated and autonomous exploration methods [21–27].

Today, automated and autonomous experiments are becoming a key component of scientific research across multiple disciplines [14,21,28]. For instance, the rapidly decreasing cost of automated synthesis tools, including pipetting and microfluidic robotics, along with capabilities for fast and high-throughput screening of traditional combinatorial libraries, and the emergence of fully automated labs, have already made a significant impact on the investigation of materials such as hybrid perovskites [29,30], battery materials [31,32], and more [33]. It is broadly anticipated that in several years synthesis robots and automated labs will be a part of materials science and chemistry laboratories across the academia and industry.

This rapid development in autonomous experimentation brings forth the next challenge – to introduce the theory into the exploration loops. Indeed, currently, the experimental workflows in automated labs are typically based on data-driven myopic workflows with Bayesian optimization (BO) on Gaussian Processes (GP) under the hood [18,28,34–36]. The theory is sometimes employed as a guide to the initial object selection[37] or may be integrated via multifidelity approaches [38–40]. At the same time, to our knowledge, the reliable framework for the key element of scientific progress – the development and update of theoretical models themselves during the experiment – has not been realized yet.

The capacity to dynamically design and refine theoretical models to accurately describe specific materials or material systems is particularly important for the developing "*digital twins*," representing a cutting-edge focus within the material science community [41]. "Digital twin" is a high-fidelity digital representation closely mirroring the form and the functional responses of a specified physical object [41]. Materials themselves embody highly intricate multiscale and multiphysics systems with a plethora of interdependent phenomena determined on length scales spanning from the atomic to the macroscale. To be effective, "digital twins" must replicate the characteristics, properties, and performance of the physical material across various length scales, containing both correlative data and causative models. This replication enables accurate prediction, optimization, and informed decision-making for material design and feature engineering.

Here, we present a *Bayesian co-navigation framework* designed to guide the exploration of material system functionality through the concurrent development of surrogate models based on theoretical calculations and experimental investigations. The core concept of co-navigation lies in the real-time tuning of theoretical model hyperparameters while simultaneously running theory and experimental discovery or optimization cycles. This dynamic adjustment aims to reduce the mismatch between theoretically calculated and experimentally measured characteristics, adapting the model to the description of the specific material system. An adapted theoretical model with the help of real experimental observables accurately captures the underlying microscopic physical phenomena and their impact on the macroscopic functionality of the material system. An adapted theoretical model in combination with correlative surrogate models can serve as a "*digital twin*" of real material systems. Importantly, such an adapted theoretical model can tackle the inverse problem and forecast the crucial microscopic material changes required to produce desired functionality through the optimization of theoretical model parameters. The concept of Bayesian co-navigation is showcased here for the exploration of ferroelectric materials functionality. However, the framework's applicability extends to a diverse array of domains, including the exploration of optical properties in nanoclusters, microstructure-dependent properties in complex materials, and properties of molecular systems.

## 2. Experimental

Realization of the co-navigation approach requires access to both active experimental and modelling cycles. Here, the efficacy and reliability of the co-navigation framework were evaluated through the examination of the local ferroelectric properties of a $PbTiO_3$ thin film. The properties were characterized using Band-Excitation Piezoresponse Switching Spectroscopy (BEPS) local hysteresis loops [42], measured by Scanning Probe Microscope (SPM) and theoretically calculated based on the domain patterns near the SPM probe using the FerroSim spin lattice model [43]. The experiment has been done in the simulation mode with the help of the pre-acquired ground truth BEPS dataset.

### 2.1 Sample

As a model sample, we utilized a $PbTiO_3$ (PTO) thin film on the $KTaO_3$ substrate with a conductive $SrRuO_3$ interlayer. The sample was fabricated by chemical vapor deposition method [44]. The ferroelectric structure of the $PbTiO_3$ film is characterized by domains exhibiting out-of-plane polarization orientation (*c-domains*) and in-plane polarization orientation (*a-domains*). The former aligns with regions with high Piezoresponse Force Microscopy (PFM) amplitude, while the latter corresponds to areas with a response close to zero (Figure 1). The alteration of *a-* and *c-domains* forms periodical striped structures.

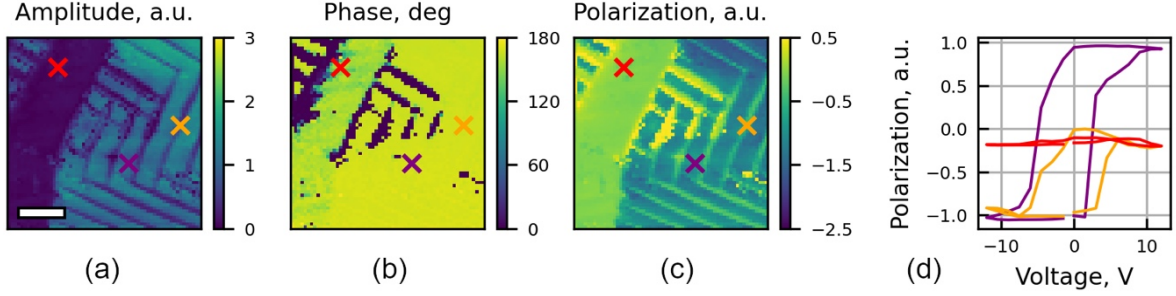

**Figure 1**. Typical domain structure of the PbTiO$_3$ film represented by (a) amplitude, (b) phase, and (c) polarization of BE PFM signal. (d) Typical local ferroelectric hysteresis loops measured in BEPS mode in the *a-domain* (red), in the *c-domain* (purple), and in the *c/a* domain wall (orange). The scale bar length is 1 µm.

*2.2 Dataset*

The study has been carried out in the simulation mode with the help of the pre-acquired dataset containing local ferroelectric hysteresis loops. This dataset was generated through local ferroelectric loop measurements in a 60x60 grid, covering a local area of 4.2x4.2 µm. The BEPS measurements were performed using Asylum Research Cypher microscope (Oxford Instrument) by coated conductive AFM probes Multi75E-G Cr/Pt (Budget Sensor) with stiffness ~3 N/m. The experimental setup includes input/output DAQ card and chassis operated by a LabView framework.

*2.3 FerroSim spin lattice model.*

To model ferroelectric film, we employ the FerroSim spin lattice model, which represents ferroelectric materials as a two-dimensional lattice comprising local polarizations $p$ at each site[45]. The local free energy at the $(i,j)$ site can be expressed in the Ginsburg-Landau-Devonshire (GLD) form:

$$F_{ij} = \frac{\alpha}{2} p_{ij}^2 + \frac{\beta}{4} p_{ij}^4 - E_{loc}(i,j)\, p_{ij}, \quad (1)$$

where $\alpha = \alpha_0(T - T_c)$ and $\beta$ are the standard GLD coefficients, $T$ – temperature, $T_c$ – Curie temperature, $E_{loc}(i,j)$ – local electrical field. The local electric field derives from the superposition of the depolarization field ($E_{dep}$), externally applied electrical field ($E_{ext}$), and intrinsic defect field ($E_d(i,j)$):

$$E_{loc}(i,j) = E_{ext} + E_{dep} + E_d(i,j) \quad (2)$$

The depolarization field is taken spatially uniform $E_{dep} = -\alpha_{dep}<p>$, where $<p>$ is mean polarization, and $\alpha_{dep}$ is a depolarization constant. The total free energy is defined as:

$$F = \sum_{ij}\left[F_{ij} + K \sum_{k,l}(p_{i,j} - p_{i+k,j+l})^2\right], \quad (3)$$

where the term $K \sum_{k,l}(p_{i,j} - p_{i+k,j+l})^2$ represents the nearest neighbor interaction with coupling constant ($K$), the sum is performed over the chosen neighborhood ($k,l$) for each site in the lattice[43].

The evolution of domain structure is defined through the Landau-Khalatnikov equation:

$$\gamma \frac{dp_{ij}}{dt} = -\frac{\partial F}{p_{ij}}, \tag{4}$$

where $\gamma$ is a dynamic coefficient, representing the speed of the domain wall mobility.

The local domain arrangement was embodied by the grid consisting of the 12x12 cells. Each cell was categorized into either the *c*-domain or the *a*-domain. The local polarizations are oriented in the *zy* plane. We applied a sinusoidal electric field that oriented along the *z*-direction, representing the out-of-plane axis in the real PTO film, with an amplitude equivalent to twice the coercive field ($E_c$). The in-plane axis was aligned with the *y*-axis.

We defined the *a*-domains as regions exhibiting significantly elevated defect electric fields, equivalent to $30E_c$, oriented in the in-plane *y* direction ($\vec{E}_d = (0, 30E_c)$). The *K* coupling constant within the *a*-domains is set to zero, ensuring the independence of these regions from polarization switching in neighboring cells. The defects filled inside the *c*-domains are taken to equal zero in both directions (Figure 2a). The application of the external electric field ($E_{ext}$) induces polarization switching of these regions and its orientation along the out-of-plane axis. We utilize the neighboring coupling constant, $K$, in cells associated with the *c*-domains as a theoretical model hyperparameter. This parameter ranges from 0 to 10 and is used to characterize the influence of neighboring domains on the local polarization within specific cells. The depolarization coefficient $\alpha_{dep} \in (0,1)$ was used as a second hyperparameter. The hysteresis loop shape undergoes substantial alterations due to variations in the hyperparameters (Figure 2b). The automated experiment aimed to minimize the discrepancy between experimentally measured and theoretically predicted hysteresis loops by adjusting the FerroSim hyperparameters $\alpha_{dep}$ and $K$.

It's important to note that the FerroSim spin-lattice model is a simplified representation and does not fully capture the complexities of real local ferroelectric structures. In particular, it lacks explicit treatment of elastic effects. Moreover, it doesn't account for the effects of local polarization switching with non-uniform field distribution and various parasitic influences. However, despite its limitations, this model yields a broad range of the possible hysteresis loop shapes dependent on domain structures and hence the capabilities of this model are assumed to be adequate for evaluating the effectiveness of the co-navigation framework.

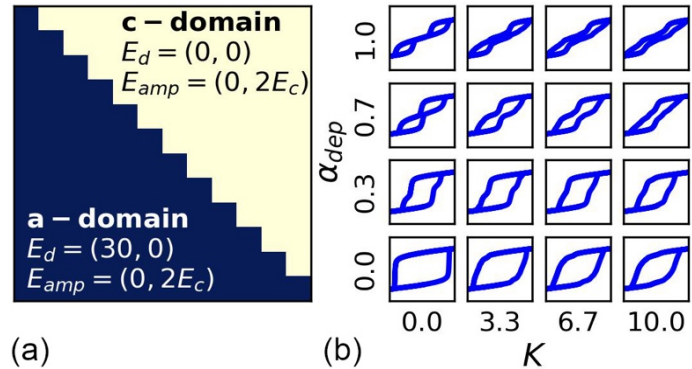

**Figure 2.** Representation of the real PTO domain pattern in the object space of the FerroSim theoretical model. (a) *a-c* domain structure and (b) correspondent hysteresis loops for various $\alpha_{dep}$ and $K$.

## 3. Co-navigation workflow: a concept

The co-navigation framework operates through two concurrent iterative exploration loops running in parallel: one theoretical (Figure 3, T-loop depicted in blue) and one experimental (Figure 3, E-loop depicted in green). These loops are independent of each other, and their objective is to explore target functionality predefined by an operator. In some cases, the object spaces for experimental and theoretical exploration loops may be identical. At the same time, the overarching requirement is the capability for theoretical and experimental object spaces to be common, albeit sampled differently for theory and experiment. The translation of the object space of a real system into a representation suitable for material discovery through a theoretical model often requires tailored, *ad hoc* solutions. This process demands finding a balance between numerous necessary approximations and simplifications on one side while ensuring that the resulting representation effectively captures the key features of the real objects. In our case of exploring ferroelectric functionality, the object spaces for both theoretical and experimental investigations are represented by the local arrangement of domains. The specific representation of real domain structures of a PTO film as objects for the FerroSim theoretical model is detailed in section 2.3.

Taking that both experimental measurements and calculation by the theoretical model are resource-intensive and time-consuming, the explorations are performed through the *surrogate* models. The surrogate models are learned on the initially explored objects to predict the mean values and corresponding uncertainties of target functionality for all objects within the object space. For spaces composed of low-dimensional objects, GP can be directly used as a surrogate model. However, for high-dimensional objects, the Deep Kernel Learning (DKL) algorithm is employed to reduce dimensionality[46,47]. The outcomes of the surrogate models are utilized to select the next object for high-cost theoretical calculations or experimental measurements using the active learning Bayesian paradigm through the maximization or minimization of acquisition functions (Maximum Uncertainty, Upper Confidence Band, etc.). After the acquisition of new data, the corresponding surrogate model is updated accordingly, and the cycle iterates.

In the co-navigation approach, the theoretical and experimental independent surrogate models may be guided by different acquisition functions and possess different rates of exploration, constrained by the available latencies of the experimental and theoretical methods, respectively. If both the E-loop and T-loop share the same predefined objective, their independent exploration results can be compared through the additional validation circuit (Figure 3, depicted in red). For ferroelectric investigation, the objectives may include discovering ferroelectric functionality within the object space or determining the optimal domain pattern to achieve the desired ferroelectric response, such as a local hysteresis loop with a specific shape or area. The effectiveness of the exploration will be evaluated by observing the reduction in average uncertainty with iteration number, as well as through the application of similar advanced forensic methods [48].

The core concept of the co-navigation framework is an active feedback system, which aligns the experimental and theoretical loops to dynamically adjust the hyperparameters of the theoretical model. The objective of tuning the theoretical model iteratively is to progressively minimize the epistemic uncertainty between theoretical model predictions and experimental measurements. To design such active feedback, we introduced the **outer theory update loop** (Figure 3, circuit depicted in violet). The object space of this loop encompasses the diverse compositions of theoretical model hyperparameters. Through iterative navigation within this object space, the loop strives to converge towards optimal hyperparameter configuration to minimize the disparity between experimental and theoretical outcomes. To estimate the epistemic uncertainty associated with distinct hyperparameter selections, objects from the space of the experimental model are utilized, incorporating both already measured and extant ones. The target functionality is predicted for these objects directly or through their counterparts (twins) in the theoretical object space. Recognizing the computational expense of theoretical calculations, a surrogate theoretical model is leveraged for estimating functionality rather than the theoretical model itself. This approach balances the effectiveness of prediction with minimizing resource consumption. In a basic scenario, the mean square error (MSE) between theoretical and experimental predictions serves as a representation of the epistemic uncertainty. Thus, the main objective of the outer theory update loop would be to minimize MSE by discovering optimal theoretical model hyperparameters. The convergence of the MSE to zero reflects the ideal scenario where the theoretical model precisely encapsulates the desired functionality of a real material system. The exploration in the outer theory update loop progresses by the Bayesian optimization algorithm steered by an acquisition function featuring distinct exploitation components (Expected Improvement, Upper Confidence Band, etc.).

The co-navigational workflow offers the ability to incorporate physical knowledge by introducing the mean functions of the GP of surrogate models or theory updates loop to expedite the functionality exploration or facilitate the adjustments of the theoretical model. Additionally, they can be complemented with any form of human-in-the-loop intervention to further accelerate exploration.

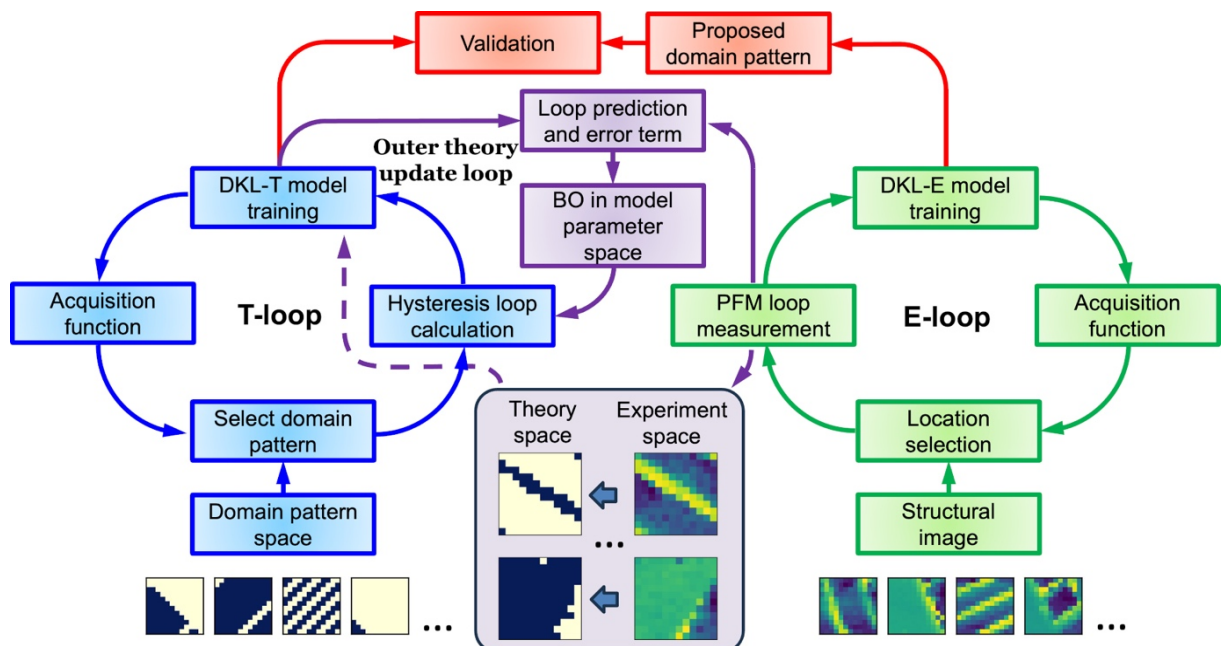

**Figure 3.** The framework for the Bayesian co-navigation of the local ferroelectric properties' exploration. The green and blue loops symbolize the concurrent exploration of experimental and theoretical object spaces through surrogate DKL models. The outer theory update loop (depicted in violet) aligns these lops to dynamically adjust the hyperparameters of the theoretical model, aiming to minimize discrepancy with experimental results. The red segment of the scheme can be employed to compare the exploration outcomes of surrogate DKL models.

## 4. Results

The efficacy of the proposed co-navigational framework has been evaluated through the simulation of autonomous experiments (AE) aimed at exploring the local ferroelectric properties of the PTO thin film. The theoretical description utilized the FerroSim spin-lattice dynamic model, while the experimental exploration was represented by local BEPS hysteresis loop measurements. The disparity between the experiment and the theoretical model was quantified through the MSE between the areas of the measured and calculated hysteresis loops. To make the experimental and theoretical data comparable the defined areas were independently normalized within the range (0,1) for MSE estimation. The resulting MSE values were multiplied by 10 before training the GP.

The theoretical object space for the FerroSim model encompasses two regions representing the *a*- and *c*-domains of a PTO film. Additionally, the *c*-domain area is intersected by the periodic structure of *a*-domain corrugations. Parametrization of the object space involves utilizing the position of the *a/c* ferroelastic domain wall, phase, and period of the corrugation structure (Figure 4a). The experimental object space consists of local domain patterns within the PFM structural image of the real PTO film (Figure 1c). We use binarization to reflect the experimental object (patch of PFM image) into the theoretical object space for prediction of the hysteresis in the outer theory updates loop (Figure 3).

The AE simulation was realized via two modes: 1) *theory-theory exploration*, and 2) *theory-experiment exploration*.

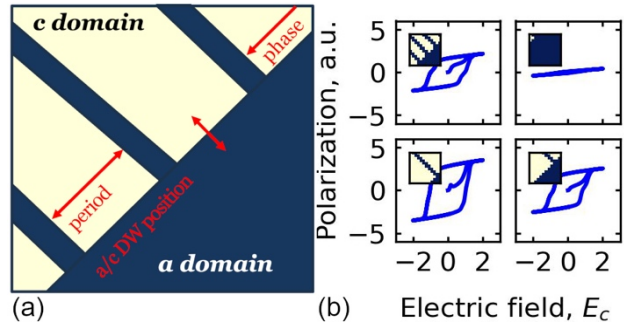

**Figure 4.** (a) Parametrization of the domain arrangements for FerroSim theoretical model object space, (b) examples of the ferroelectric loops calculated for different domain patterns.

*4.1 Theory-theory AE*

At the zero point, we assess the efficiency of the proposed co-navigation framework by the simulation of the experiment in *theory-theory* mode. Here, we substitute the experimental local hysteresis loop measurements with theoretical calculations using the FerroSim model, possessing predefined fixed values for $a_{dep}$ and $K$. Concurrently, the theoretical exploration is conducted using another FerroSim model, that exploits $a_{dep}$ and $K$ in the role of hyperparameters. The objective of the outer theory update loop is to minimize the MSE between the areas of the calculated hysteresis loops. This involves discovering the fixed values of $\alpha_{dep}$ and $K$ for the 'experimental' FerroSim model. The AE simulations in the theory-theory mode have been conducted both with and without the inclusion of the specific mean function in the GP of the outer theory update loop.

*4.1.1. Surrogate DKL models*

We employed surrogate DKL models in both T- and E-loops, utilizing the Maximum Uncertainty (MU) acquisition function. The theoretical and experimental investigations progressed with equal exploration rates. The simulations involved 600 exploration steps, and at every 20th exploration step, the theoretical model hyperparameters $\alpha_{dep}$ and $K$ were updated by the outer theory loop. The couples of $(K, \alpha_{dep})$, where $K \in (0,10)$ and $\alpha_{dep} \in (0,1)$ form the object space for the theory update loop. After the initial three random selections of the hyperparameters couple, the loop transitioned into the Bayesian Optimization paradigm driven by the Upper Confidence Bound acquisition function (UCB). The objective of the third loop is to minimize MSE between the areas of experimentally measured hysteresis loops and those predicted by the surrogate theoretical model on the objects reflected from real domain local arrangement from the experiment.

The progress of the investigation by E- and T-loops was characterized by learning curves – dependencies of the prediction uncertainty averaged throughout the object space on the exploration step (Figure 5a,c). The consistent and gradual decrease in the mean uncertainty observed during the experimental exploration signifies the natural evolution of the exploration process guided by the MU acquisition function (Figure 5a). The same background trend is obtained for the theoretical DKL model as well. However, the learning curve of the T-loop exploration exhibits explicit sharp deviations from the main trend and fluctuations in multiple regions throughout the exploration process (Figure 5c). This phenomenon is attributed to the partial inconsistency of the previously acquired data resulting from the multiple adjustments made to the theoretical model hyperparameters during the exploration. To mitigate the influence of data acquired with other hyperparameters, we constrained the training dataset at each step to include only the last discovered 20 locations. It is important to note that with such a limitation, the side impact is confined solely to the previous hyperparameter set, gradually diminishing from 100% immediately after the theoretical model updates to zero until the next update cycle. The surrogate theoretical DKL models employed for predictions in the outer theory update loop are trained only by the data acquired for the actual hyperparameters. The constraints on the learning dataset for the theoretical model can be lifted once the outer updates loop converges to the optimal hyperparameters, signifying the adaptation of the theoretical model to accurately describe the specific real material system.

The introduction of a mean function with a specific form into the outer theory update loop does not affect the investigation conducted by the surrogate models. The curves presented in Figure 5 correspond to the experiments conducted using the outer theory update loop without the introduction of the special mean function into the GP. Meanwhile, the results obtained for the GP with such an introduction are showcased in the Supplementary Materials (Figure S1).

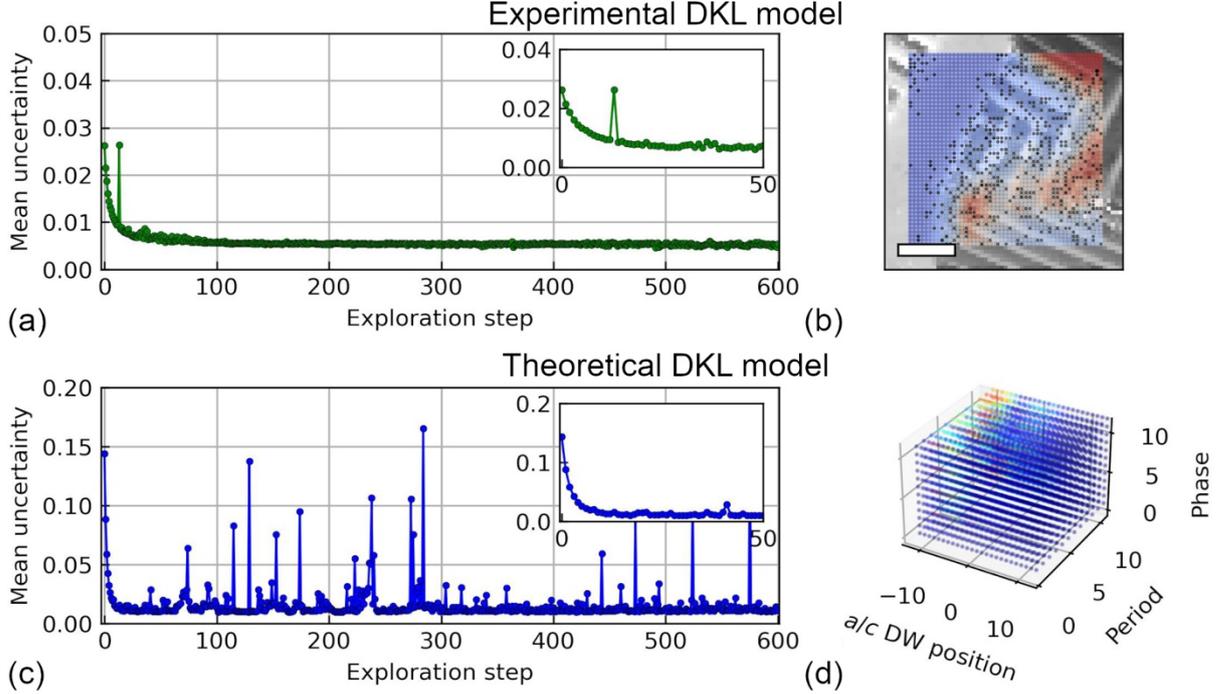

**Figure 5.** Exploration by surrogate DKL models: average GP uncertainty for the (a) experimental and (c) theoretical models predictions of hysteresis loops area within the object space. The initial exploration steps are duplicated to the insets. Latent spaces in which the (b) experimental and (d) theoretical surrogate models operate, with values of the hysteresis loop areas predicted by the models at the last exploration step depicted by color. The scale bar length is 1 μm.

*4.1.2. Outer theory update loop*

As mentioned earlier, we conducted experiments in two modes based on the parameters of the outer theory update loop: one without introducing a specific mean function into the outer theory update loop (standard GP), and the other with such an introduction (structural GP). It is important to note that even in the case of a standard GP without a specific form of a mean function, the mean function has been slightly modified. In the classical GP algorithm, the zero function is typically used as a mean function. However, in the outer theory update loop, the explored MSE is non-negative, which results in the prioritization of locations at the latent space edges, when dealing with a zero predefined mean function. While the prioritization of field edges is less critical in the 1D exploration space, where borders are confined to only two locations, it becomes indispensable for spaces with higher dimensions. This leads to a concentration of over 90 percent of chosen by algorithm locations directly on the space borders. To mitigate this effect, we have opted for the GP mean function to be equivalent to the maximum of the acquired MSE at each specific exploration step.

In the case of structural GP, we introduced the mean function in the form of the elliptical equation $\alpha_0(x - x_0)^2 + \alpha_1(y - y_0)^2$. The elliptical shape of the mean function may not perfectly align with the ground truth, but it effectively prioritizes the central region of the exploration space. Additionally, this contributes to the limitation of promptly altering hyperparameter values, meaning that $\alpha_{dep}$ and $K$ undergo only limited changes within a single update. This enhances the compatibility of data acquired with one hyperparameter set with that obtained using a subsequent set, which, in turn, improves the applicability of the surrogate theoretical model.

We observed in both experiments with standard (Figure 6a,c) and structural GP (Figure 6b,d) that the algorithms did not always precisely converge to the predefined ground truth hyperparameter set ($K = 3$, $\alpha_{dep} = 0.5$). In general, the existent mismatch with ground truth after the algorithm's convergence could be aligned to three potential reasons:
1. Insufficient exploration of the object space by the outer theory model, causes the algorithm to converge to a local minimum.
2. Inadequate training of the DKL surrogate model used to predict theoretical values of the hysteresis loops area.
3. Existence of multiple minima within the object space of outer theory update loops.

The exploration of the object space potentially could be enhanced by adjusting the $\beta$ coefficient. It was mentioned above that the outer theory update loop is driven by the minimization UCB calculated for MSE. In this case, the UCB acquisition function is defined as:

$$UCB(x) = \mu(x) - \beta\sigma(x), \tag{5}$$

where $x$ – is the location in the explored space, $\mu$ – mean value of MSE, $\sigma$ – uncertainty of MSE. The coefficient $\beta$ delineates the trade-off between exploration and exploitation within the acquisition function. The regulation of $\beta$ enables the operator to influence the evolution of the experiment. We introduced a gradual change in $\beta$ from $10^2$ to $10^{-3}$ during the exploration phase using a sigmoid function (Figure 6c,d). Boosting the exploration component at the beginning of the AE enables a more thorough exploration of the object space, enhancing the likelihood of discovering a true minimum of MSE. Simultaneously, promoting the exploitation component in the later stages of the experiment is crucial for the convergence of hyperparameter adjustment. It is evident that when $\beta$ is sufficiently large, the outer theory update loop may significantly change the model hyperparameters at each iteration (Figure 6c). Oppositely, as $\beta$ diminishes, the algorithm typically settles on the optimal hyperparameter set, indicating convergence to the minimum (Figure 6c). In some cases, we also observed residual switching between a few of the most promising hyperparameter regions, even with the predominance of the exploitation component (Figure 6d).

The challenge of insufficient dataset size for training the theoretical model is particularly significant. The requisite length of data is inherently dependent on both the complexity of the model and the diversity of theoretical objects to predict. The consistent decrease in uncertainty at the beginning of the learning curve (Figure 5c) suggests that the trained surrogate theoretical model possesses the capability to capture the intricacies within the explored theoretical object space. At the same time, increasing the length of the training dataset while reducing the frequency of updates to the theoretical model hyperparameters results in enhanced performance of the DKL surrogate model. However, this approach also slows down the overall speed of co-navigational investigation.

We do believe that the main reason for the partial convergence of the model is the existence of the valleys of the MSE minimum. It is highly resource-intensive to compute the ground truth of the MSE distribution across the hyperparameters object space for all domain patterns (theoretical objects). However, analyzing the ground truth for several domain patterns reveals that the error minimum forms contours whose size and shape vary depending on the specific arrangement of the domains (Figure S2).

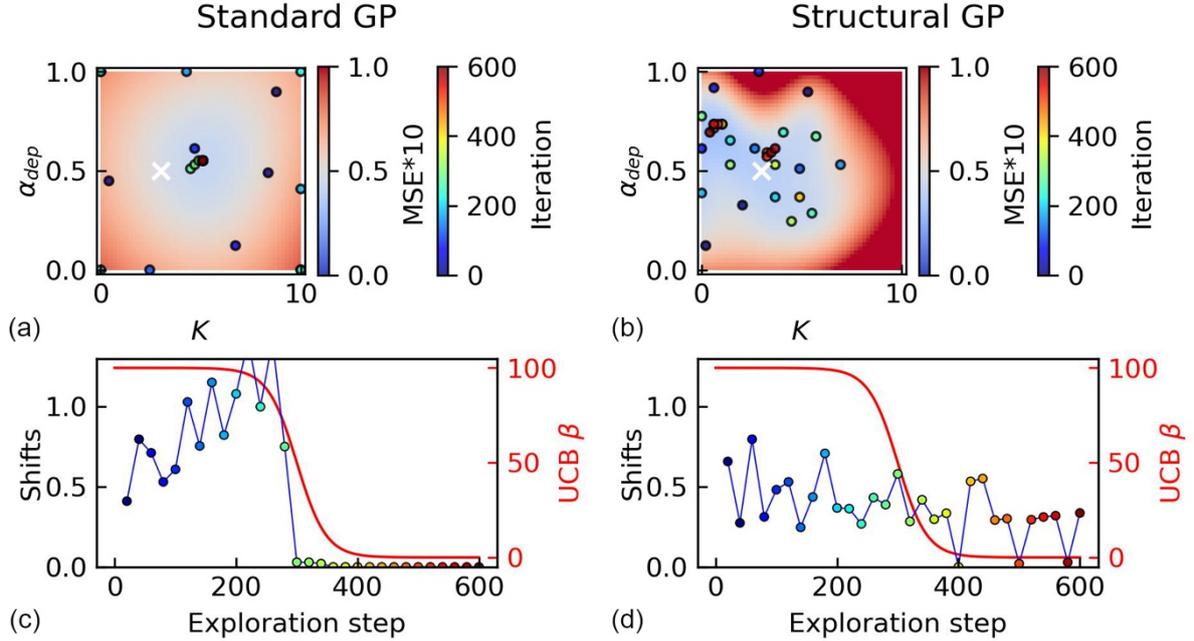

**Figure 6.** The theoretical model hyperparameter updates during the AE with (a,c) standard and (b,d) structural GP in the outer theory update loop. (a,b) Explored locations within the object space. The background color corresponds to the GP reconstruction of the MSE distribution generated by the model trained at the final exploration step. (c,d) Shifts in the explored space after the hyperparameter updates during the experiment. The range of $K$ has been normalized to $(0,1)$ to make $K$ changes compatible with $\alpha_{dep}$ changes.

We may distinguish some differences and similarities in the evolution of the AE with standard and structural GP in the outer theory update loops. It can be observed that while the model with standard GP explores the entire latent space when $\beta$ is sufficiently large (Figure 6a), the exploration of the structural GP is concentrated in a distinct region (Figure 6b). The regional prioritization depends on the seed random locations choice and the specific shape of the mean function. In the case of structural GP, following the transition to exploitation, we observe the selection of locations in the vicinity of two poles with permanent switching between them (Figure 6b,d), whereas the standard GP converges to a single minimum (Figure 6a,c). In general, both algorithms remain near the minimums discovered during the exploration stage of the experiment.

For the model utilizing standard GP, when the exploration component dominates, we observe high variability in the MSE values, whereas in the terminal stage of the experiment algorithm converges to smaller MSE values (Figure 7a). The residual variations of the MSE at the ending part are associated with the variability of theoretical surrogate models predictions. The models on different steps are trained on distinct datasets comprising different domain patterns from the theoretical object space. Interestingly, we did not observe a decrease in the MSE values with an experiment ongoing for a model with structural GP (Figure 7b). This happens because the variations in MSE for predictions made by different models (meaning models trained at different steps with different datasets) exceed the variability of MSE for different hyperparameter sets in the explored region.

The GP parameters demonstrate a similar evolution with ongoing experiments. In both cases, we observed the gradual stabilization of the kernel lengths and noise level (Figure 7c-f). The lower absolute values observed for structural GP are associated with the smaller explored region within the hyperparameter space. The downward trend for the kernel scale for the structural GP (Figure 7h) signifies a weak difference between the MSE values within the exploring space which also impacts the partial convergence of the model to the ground truth hyperparameters values.

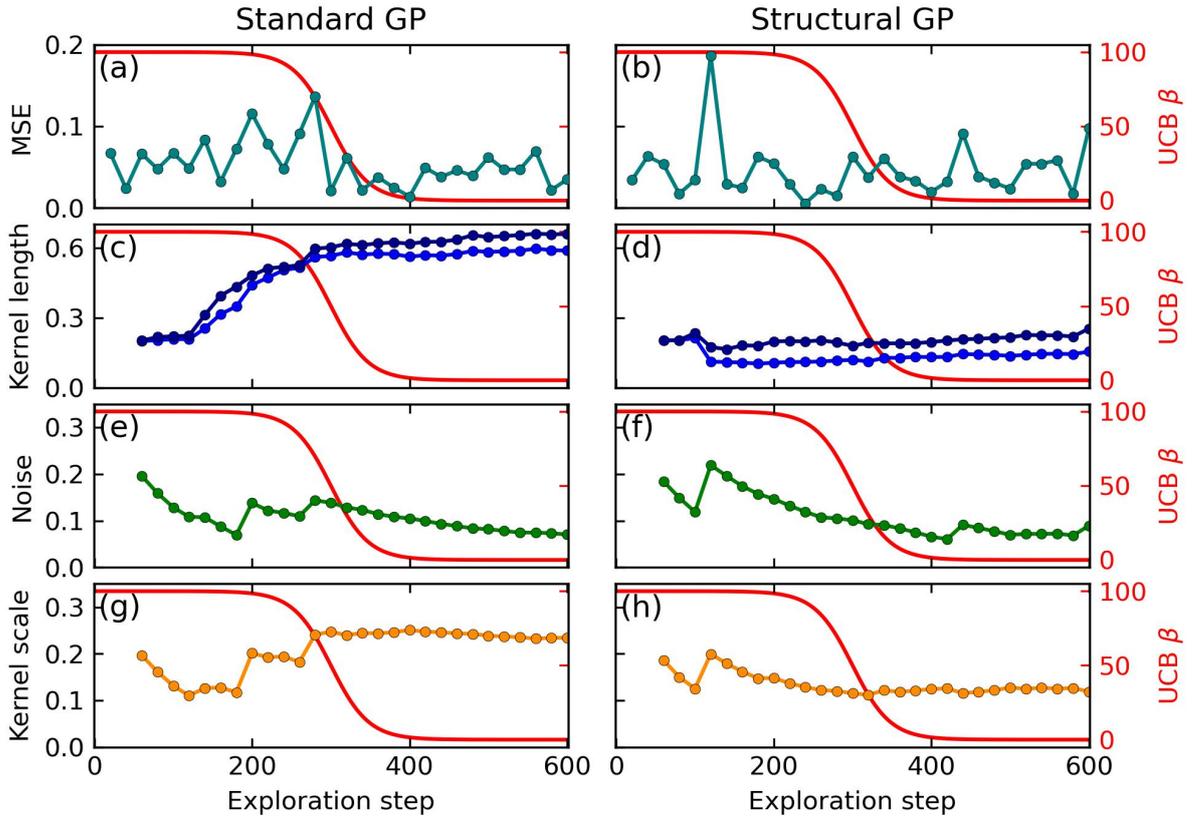

**Figure 7.** Outer theory updates loop (a,c,e,g) standard and (b,d,f,h) structural GP parameters evolution throughout the experiments: (a,b) MSE, (c,d) kernel lengths of latent processes, (e,f) noise, (g,h) kernel scale. The red graphs represent the changing of the UCB $\beta$ coefficient with the exploration step.

*4.2 Theory-experiment AE*

To simulate conditions closer to those of a real experiment, the effectiveness of the proposed framework was evaluated through simulation in a *theory-experiment* mode. Here, real measured hysteresis loops were utilized to train the experimental surrogate model. The simulation also consisted of 600 exploration steps with FerroSim hyperparameters updates every 20 steps. We exploited the outer theory update model with standard GP. The priors, $\beta$ evolution, and experiment protocol were the same as those performed in the *theory-theory* simulation.

The learning curves of surrogate models from E- and T-loops show similar behavior as in the theory-theory simulations (Figure 8a,b). The evolution of the GP hyperparameters (Figure 8f-h) was the same as in the theory-theory experiment. After exploring the latent space of hyperparameters, the model converges to a single region (Figure 8c,d), accompanied by a noticeable decrease in MSE values (Figure 8e).

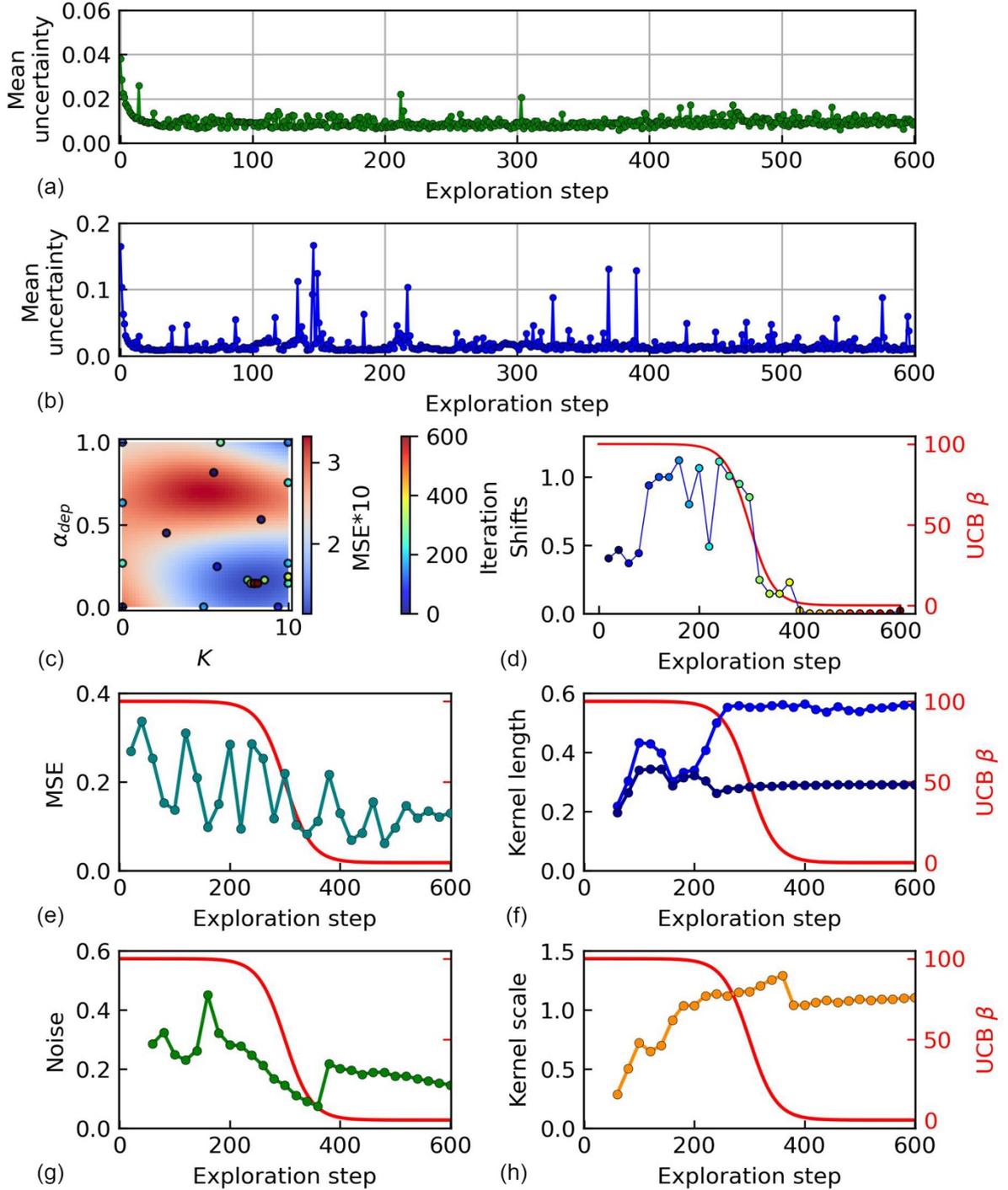

**Figure 8.** Autonomous experiment in theory-experiment regime: learning curves of (a) experimental and (b) theoretical DKL surrogate models; (c) exploration trajectory in the hyperparameter space of the theoretical model and (d) correspondent shifts after the theory hyperparameters updates. To determine shifts the range of *K* has been normalized to (0,1). Dependencies on the exploration steps of (e) MSE, (f) GP kernel lengths, (g) GP noise, and (h) GP kernel scale.

## 5. Discussion

The observed dynamic of the MSE error during the theory-experiment simulation confirms the robustness and efficiency of the proposed co-navigational framework in decreasing the epistemic uncertainty between the theoretical model and experiment. The outcome of the co-navigation results lies in a multi-component system of models. This system integrates experimentally measured data, a surrogate experimental model used to predict target functionality within the specific region, a fine-tuned theoretical model for a precise description of the functionality of a specific material system, and a surrogate theoretical model. The surrogate theoretical model, that capable of predicting target functionality and related uncertainty, plays the role of the fast low-cost counterpart of a resource-intensive theoretical model. The performance of theoretical surrogate models holds the potential for infinite enhancement with the availability of newly acquired experimental data or more advanced theoretical models. Access to both a precise theoretical model and its fast low-fidelity counterpart significantly enhances the flexibility of the resulting system of models and broadens the potential range of its applications. The main framework's strength lies in the dynamic improvement of the listed components, which occurs without necessarily requiring human intervention in the loop. Furthermore, co-navigation enables a broad spectrum of human interventions, including dynamic switching between acquisition functions, real-time adjustment of the $\beta$ coefficient of the UCB in the outer theory update loop, incorporation of predefined physical knowledge in the form of mean functions, modification of the training dataset length for the theory surrogate model, and more.

Refined through co-navigation, the theoretical model evolves into a *dynamic digital twin* of the specific material system [41]. This twin not only offers insights into the current state and potential future behavior of the specific material system but also unveils underlying physical mechanisms. Important to repeat, that the FerroSim lattice model employed in this study for the co-navigation demonstration serves as a simplified representation. This "toy" model does not encompass the vast array of interrelated phenomena that influence local ferroelectric behavior. At the same time, the proposed framework can be applied to any type of model without imposing specific requirements regarding their complexity or nature. The real theoretical model may incorporate various parameters, including material properties, environmental factors, and any other relevant variables that affect the behavior of the material. In the case of comprehensive and complex models encompassing various system properties, the implementation of multi-objective optimization strategies, such as multi-objective BO, within the outer theory update loop, can aid in achieving a balance among accurately describing multiple target properties. Furthermore, the co-navigational framework can operate inversely, facilitating the optimization of experimental investigation based on theoretical feedback. For instance, in ferroelectric domain engineering, we can dynamically adjust parameters such as the applied pressure force to the tip, the amplitude, duration, and shape of the applied voltage waveform, and more to design a specific domain arrangement based on the theoretical prediction of the target functionality.

## 5. Summary

In conclusion, we have presented an efficient co-navigational workflow tailored to guide the exploration of material system functionality by theoretical model and experimentally with simultaneous calibration of the theoretical model transforming it into the digital twin of the investigating material. The framework operates through two concurrent and independent exploration cycles driven by Bayesian optimization: one theoretical and one experimental. These cycles are aligned with each other through the third outer theory update loop. This loop aims to reduce the epistemic uncertainty between the theoretical model's calculations and experimental outcomes by fine-tuning the theoretical model to accurately represent the specific material system. As a result of the iterative optimization, the calibrated theoretical model evolves into a digital twin of the material system. This digital twin possesses the capability not only to depict the system's behavior accurately but also to uncover its underlying physical mechanisms.

The effectiveness of the proposed frameworks has been validated by the investigation of the local ferroelectric properties of the $PbTiO_3$ thin film. The theoretical model was represented by the FerroSim lattice spin model, which predicted polarization hysteresis by accounting for local domain arrangements. The local ferroelectric loops were experimentally measured using a scanning probe microscope via Band Excitation Switching Spectroscopy. In the performed experiments we observed steady trends in the minimization of the mismatch between experimental and theoretical calculation that confirms the reliability of the co-navigational approach. We introduced and discussed various modifications and adjustments of the GP BO, such as implementing advanced mean functions and on-the-fly managing the exploration/exploitation ratio via UCB $\beta$ coefficient, to accelerate the convergence of the theoretical mode optimization.

The implementation of the co-navigational approach is expected to significantly simplify the creation of digital twins for the materials. Indeed, co-navigation proposed a powerful way for real-time theoretical model calibration based on up-to-date experimental feedback in automated or semi-automated regimes. The co-navigational approach can be used for a diverse array of systems and theoretical models without any limitation on the nature of describing phenomena or model complexity, where the theoretical exploration and experimental measurements can be performed in the identical or reflecting to each other object spaces and their outputs are comparable. The potential applications list includes but is not limited to such as molecular discovery, plasmonic assembly, and more. We believe that a co-navigation framework may revolutionize the production of digital twins and approaches to material modeling in general.

**Acknowledgments**

The development of Bayesian co-navigation workflow was supported (SVK) by the UTK start-up funding. The experimental measurements (YL) and development of FerroSim library (RKV) were supported by the Center for Nanophase Materials Sciences (CNMS), which is a US

Department of Energy, Office of Science User Facility at Oak Ridge National Laboratory. The development of GPax Python package was supported (MAZ) by the PNNL LDRD project. This work was partly supported by MEXT Initiative to Establish Next-generation Novel Integrated Circuits Centers (X-NICS) (JPJ011438) and Data Creation and Utilization Type Material Research and Development Project Grant Number JPMXP1122683430 (H. F.).

**Author Contributions**

**Boris N. Slautin**: Conceptualization (equal); Software (equal); Data curation (lead); Writing – original draft. **Yongtao Liu**: Data curation (equal); Writing – review & editing (equal). **Hiroshi Funakubo**: Resources. **Rama K. Vasudevan**: Software (equal); Data curation (equal); Writing – review & editing (equal). **Maxim A. Ziatdinov**: Software (equal); Writing – review & editing (equal). **Sergei V. Kalinin**: Conceptualization (lead); Methodology; Supervision; Writing – review & editing (lead).

**Conflict of Interest**

The authors have no conflicts to disclose.

**Data availability**

The code that supports the funding is publicly available at https://github.com/Slautin/2024_Co-navigation/tree/main. The FerroSim codes are available at https://github.com/ramav87/FerroSim. The GP codes are implemented using GPax package https://github.com/ziatdinovmax/gpax.

**Bayesian Co-navigation: Dynamic Designing of the Materials Digital Twins via Active Learning**

Boris N. Slautin, Yongtao Liu, Hiroshi Funakubo, Rama K. Vasudevan[2], Maxim A. Ziatdinov[4], and Sergei V. Kalinin

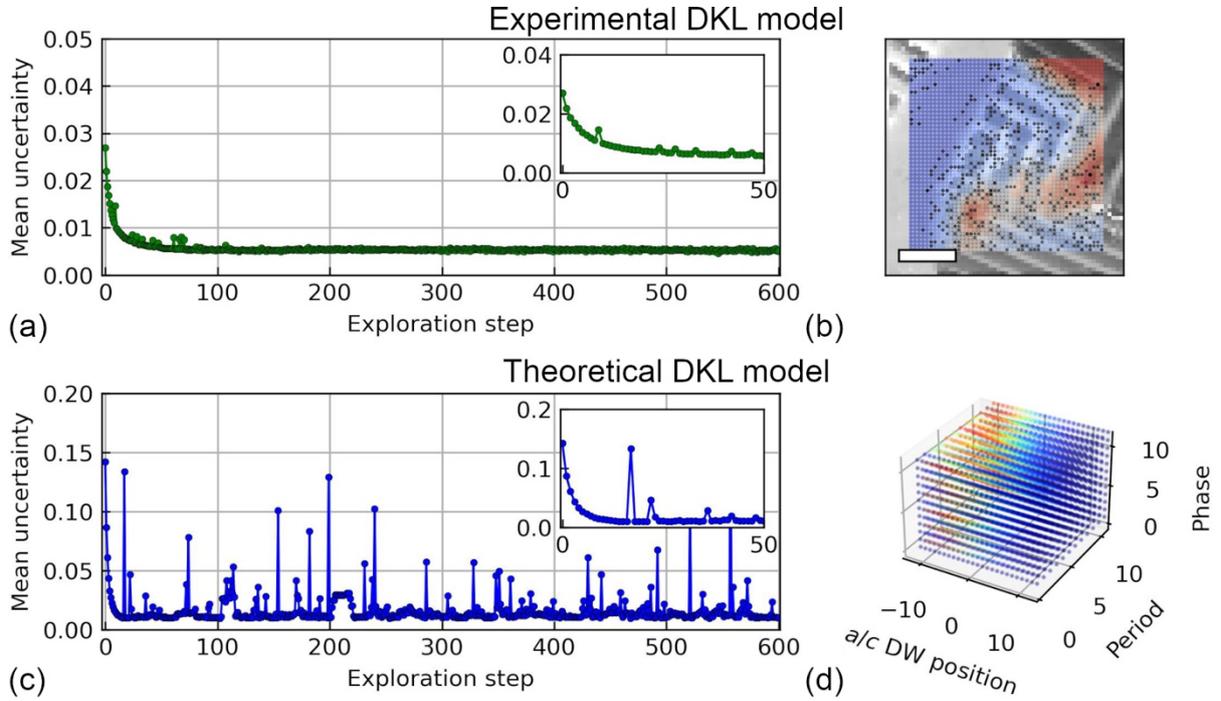

**Figure S1.** Surrogate DKL models investigations for experiment with outer theory update loop with **structural GP**. The average GP uncertainties for the (a) experimental and (c) theoretical models predictions of hysteresis loops area. The initial exploration steps are duplicated to the insets. Latent spaces in which the (b) experimental and (d) theoretical surrogate models operate, with values of the hysteresis loop areas predicted by the models at the last exploration step depicted by background color. The scale bar length is 1 μm.

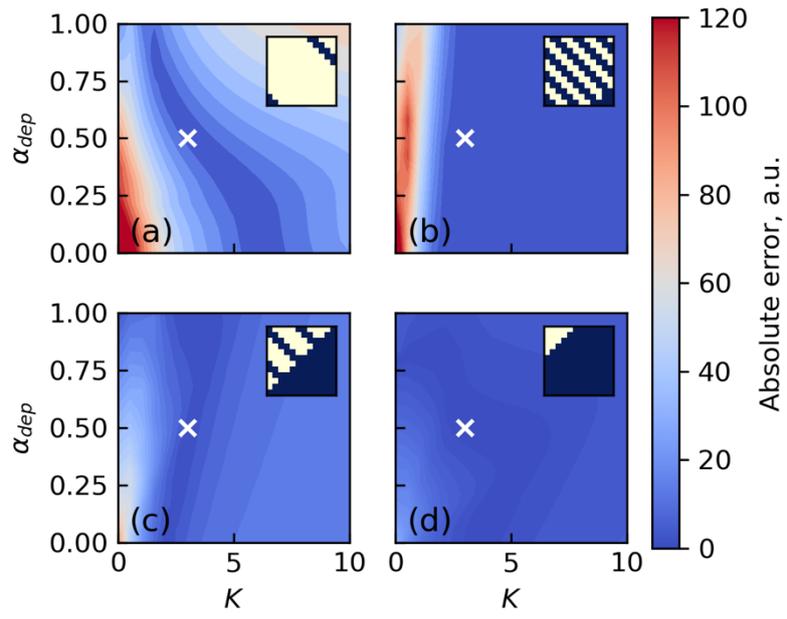

**Figure S2.** Ground truth absolute error distributions for the different domain patterns.